\documentclass[preprint,showkeys,showpacs,preprintnumbers,amsmath,amssymb]{revtex4}

\usepackage{graphicx}
\usepackage{dcolumn}
\usepackage{bm}
\makeatletter
    \newcommand{\rmnum}[1]{\romannumeral #1}
    \newcommand{\Rmnum}[1]{\expandafter\@slowromancap\romannumeral #1@}
    \makeatother

\begin{document}
\preprint{arXiv:0706.1338}

\title{Supersymmetric Hamiltonian Approach to Edge Excitations  in $\nu = 5/2$  Fractional Quantum Hall Effect}

\author{Ming Yu}
 \email{yum@itp.ac.cn}
\author{Xin Zhang}
 \email{xzhang@itp.ac.cn}
\affiliation{Institute of Theoretical Physics, CAS, P.O.Box \rm{2735}, Beijing 100080, P.R.China}
\date{Sep. 10, 2007, revised}

\begin{abstract}
A supersymmetric Hamiltonian is constructed for the edge excitations of the Moore-Read
(Pfaffian) like state, which is a realization of the N=2 supersymmetric CS model.
Fermionic generators and their conjugates are introduced to deal with the fermion pairing,
whose condensation form a BCS like state.  After
Bogoliubov transformation, a N=2 supersymmetric and
nonrelativistic Hamiltonian is found to take a known form, which is integrable.
The main difference between the Moore-Read state
and our BCS like state is that the number of fermion pairs in our
formalism is not fixed. However, we have also found
that the excited states in our model looks similar but not exactly
the same as Moore and Read's.
\end{abstract}

\maketitle

\section{Introduction}

In recent years, Fractional Quantum Hall Effect(FQHE) has attracted
renewed interests because of its promising applications in
topological quantum
computing\cite{nayak}\cite{sarma}\cite{stern}\cite{bonderson}\cite{hou}\cite{yueyu1}.
The $\nu = 5/2$ quantum Hall system is particularly interesting
because of its nonabelian
statistics\cite{wu}\cite{zhang}\cite{jain}\cite{wen1}\cite{haldane}\cite{read1}. Since the braiding
is topological in nature, we expect a more stable interwining
relation against quantum decoherence. Though the 2d quantum Hall
system is not exactly solved, nevertheless, Laughlin's wave function
provides a very good approximation to the true ground state of the
fractionally filled quantum Hall system with filling factor $\nu =
1/m$, $m$ odd\cite{laugh}. The case of $\nu = 1/m$, $m$ even,
however, is a different story. The most interesting case  is the one
with $\nu = 5/2$, in which the groundstate can be approximately
described by the Moore-Read (Pfaffian) wave function\cite{read1}.
The quasi excitations in FQHE are  quasi-holes, whose creation  are
energy costly in the bulk of the 2d plane but on the boundary, where
there exists gapless edge excitations. In this sense, fractional
quantum Hall state is incompressible.

Since edge states are intimately related to the bulk properties of
the FQHE, they deserve a better and more thorough understanding. The
edge excitations for $\nu = 1/m$, $m$ odd, are known to be described
by the Calogero-Sutherland Model (CS model)\cite{cal1}\cite{suther2}\cite{suther3}, which is a
one-dimensional (1d) many-body system with long range interaction.
The enormousness of the references on CS model prevents us from
giving a thorough list of references  but a few review
articles\cite{ha}, from which more relevant papers can be found. The
edge excitations in $\nu = 5/2$ FQHE have been studied in
detail in ref.\cite{read1}. However, a Hamiltonian description, and
therefore a systematic quantum mechanical approach, is still missing
in the literature. There are also supersymmetric extensions of the
CS model\cite{mathieu1}. But none of them made direct
connection to the $\nu = 5/2$ FQHE. The main difficulty lies in the
fact that the vacuum state of the supersymmetric CS model does not
contain any Pfaffian like state.

The present work is motivated by constructing a Hamiltonian for the
edge excitations of the Moore-Read (Pfaffian) like state. We succeed
in doing so by making a direct connection between the N=2
supersymmetric CS model and the Moore-Read (Pfaffian) like state.
The construction is made by the following observations. First, there
are two kinds of distinct edge excitations. One corresponds to
charge density fluctuations, which is bosonic. The other is
fermionic, which destroy or create a fermion. Second, we consider a
condensation of fermion pairs, which form a BCS like state, as our
ground state. Third, we perform a Bogoliubov
transformation\cite{bogoliubov}, and define a new set of orthonormal
fermionic generators and annihilators, the later of which annihilate
the BCS vacuum. Finally, we construct a N=2 supersymmetric and
nonrelativistic Hamiltonian, which,  in terms of Bogoliubov
transformed variables, is exactly the same form presented in the
literature\cite{khare}\cite{mathieu1}, in which one can find that this 
system is integrable. From the
strategy mentioned above, we can see the main difference between
Moore-Read Pfaffian state and our BCS like state is that we have to
introduce fermionic variables explicitly, and the number of
fermion pairs in our ground state is not fixed. The differences
should be irrelevant in the thermodynamic limit, just like  the ones
between canonical ensemble and grand canonical ensemble. However, we have also found
that the excited states in our model looks similar but not exactly
the same as Moore and Read's.

It is never over-emphasized that the bulk theory and the edge
excitations are dual to each other. While the electrons in the bulk
are strongly correlated, the edge excitations are essentially free
boson and fermion conformal filed theory (CFT) in some cases\cite{read1}.
This can be seen by considering the ground state wave function for the
$\nu =1/m$, $m$ odd, and $\nu = 5/2$, separately. The formal is just the
chiral part of the N vertex correlation function of the c=1 free
boson conformal field theory (CFT). The later corresponds to c=3/2
free boson and free fermion superconformal field theory. Is there
any holographic principle there? The answer seems to be positive if
we take thermodynamic limit here. But we would not elaborate further
on this point. Instead, we shall show in our forthcoming paper\cite{yz} that
the Hamiltonian can also be written in the second quantized form,
which, in the thermodynamical limit, become exactly the zero mode of
the Virasoro algebra, or Super Virasoro algebra, respectively.
All the edge excitations are also expressible in terms of
free boson and fermion oscillators.

The present paper is organized as following. In sec.\Rmnum{2} we review some
basic properties of the Calogero-Sutherland model and it's relation to
the edge excitations in $\nu=\frac{1}{m}$($m$ odd) FQHE. Sec.\Rmnum{3}
is a brief introduction to the N=2 SUSY extended CS model. In sec.\Rmnum{4},
we generalize Moore-Read state to super-space formulation. In sec.\Rmnum{5}
we define the Bogoliubov transformation on fermion variables. The N=2 SUSY
extended CS model is realized with respect to the BCS like ground state.
Finally, in sec.\Rmnum{6}, we make our conclusions and some speculations
about the holographic principle.

\section{Calogero-Sutherland Model}
The Hamiltonian for the CS model\cite{suther1}\cite{suther2}\cite{suther3} is,

\begin{equation}
\mathcal{H}=-\sum^N_{i=1}\frac{\partial^{2}}{\partial
x^{2}_{i}}+\left(\frac{\pi}{L}\right)^2\sum_{i<j}\frac{2\lambda(\lambda-1)}{d
^2 (x_{i}-x_{j})}\label{h1}
\end{equation}
where we have assumed the periodic boundary condition, $x_i\equiv
x_i +L$. That is to say, $\{\frac{2\pi}{L}x_i\}$ are compactified on
the unit circle. $d(x)=|\sin(\pi x/L)|$ is the proper distance
between two points on the unit circle (the length of the straight
line connecting the two points) if their coordinates differ by $x$.
$d^2(x)$ can also be written as an infinite sum\cite{yang},
\begin{equation}
d^2(x)=\left(\frac{\pi}{L}\right)^2 \sum_{n \in Z} \frac{1}{x+nL}
\end{equation}

The ground state wave function is given by
\begin{equation}\label{psi}
\Psi_{0}=\prod_{i<j}\sin^{\lambda}[\frac{\pi}{L}(x_{i}-x_{j})]
\end{equation}
which can also be written in the $z$ coordinates  as
\begin{equation}
\Psi_{0}=\prod_{i<j}(z_{i}-z_{j})^{\lambda}\prod_{k}z_{k}^{J_{0}}
\end{equation}
where $J_{0}=-(\lambda/2) (N-1)$ and $z_{i}=\exp(i\frac{2\pi}{L}x_{i})$

All the excited eigen-states of the Hamiltonian $\mathcal{H}$ can be
given by,
\begin{equation}
\Psi=J(z_{1},z_{2},\cdots,z_{N};\lambda)\Psi_{0}
\end{equation}
where $J$ are called the Jack symmetric polynomials (symmetric polynomials in $z_{i}$'s)\cite{jack} defined
by the eigenfunction equation, $\mathcal{H}_{\lambda} J = EJ$

Here,
\begin{equation}
\mathcal{H}_{\lambda}=\sum^{N}_{i=1}(z_{i}\partial_{z_{i}})^2+\frac{\lambda}{2}\sum_{i\neq
j}
\frac{z_{i}+z_{j}}{z_{i}-z_{j}}(z_{i}\partial_{z_{i}}-z_{j}\partial_{z_{j}})
\end{equation}
To get a simple idea of how the CS states look like, we just write down the first a few low lying
excited states,
\begin{eqnarray}\label{jpsi}
  \Psi_{(1,0,0)} &=& a_1\Psi_0 ,\nonumber\\
  \Psi_{(1,1,0)} &=& (a_1^2+\frac{1}{\lambda}a_2)\Psi_0,\nonumber \\
  \Psi_{(2,0,0)} &=& (a_1^2-a_2)\Psi_0,\nonumber\\
  \Psi_{(1,1,1)} &=& (a_1^3+\frac{3}{\lambda}a_1a_2+\frac{2}{\lambda^2}a_3)\Psi_0 ,\\
  \Psi_{(2,1,0)} &=& (a_1^3+\frac{\lambda-1}{\lambda}a_1a_2+\frac{-1}{\lambda}a_3)\Psi_0,\nonumber \\
  \Psi_{(3,0,0)} &=& (a_1^3-3a_1a_2+2a_3)\Psi_0.\nonumber
\end{eqnarray}
where $a_n=\sum_{i=1}^{N}{z_i^n}$.

The example states given above are just a few examples of the Jack polynomials, which are generated by all the symmetric
polynomials in $z_i$'s. Notice that all the exited states are made of Jack polynomials generated by the products
of $a_n\equiv \sum_i z_i^n$, which can be regarded as the Fourier components of the density fluctuations.
To see how they are related to the edge excitations in $\nu = 1/m$, $m$ odd FQHE, let us consider
Laughlin's quasihole operator defined on the disk geometry,

\begin{equation}
\prod_{i=1}^N (z_i-w)=\sum_{n=0}^N f_n (-w)^n
\end{equation}

where $f_n$ are some symmetric polynomials. If $w$, the position of the quasihole, goes to infinity, the dominant
contributions come from the first low lying $f_n$'s. Thus we see that indeed the Jack polynomials, which are symmetric
polynomials, represent the gapless edge excitations.

For later analysis it is more convenient to rewrite the Hamiltonian, eq.(\ref{h1}), in the following form,
\begin{equation}\label{h2}
\mathcal{H}=(-i\partial_{x_{i}}-\frac{i\pi\lambda}{L}\cot x_{ij})(-i\partial_{x_{i}}+\frac{i\pi\lambda}{L}\cot x_{ij})+E_0
\end{equation}
where $x_{ij}\equiv \frac{\pi}{L}(x_i-x_j)$ and $E_0=(\frac{\pi\lambda}{L})^2(\frac{N(N-1)(N-2)}{3}+1)$ is the
ground state energy, hence will be omitted in the rest part of our paper without further noticing.
The equivalence between eq.(\ref{h1})  and eq.(\ref{h2}) can be checked thanks to the identity
\begin{equation}
\sum_{i\neq j\neq l}\cot x_{ij}\cot x_{il}=-\frac{1}{3}N(N-1)(N-2)
\end{equation}

One of the advantage for writing Hamiltonian in the form of eq.(\ref{h2}) is that the ground state wave
function $\Psi_{0}$ is annihilated by $\mathcal{A}_i$
\begin{equation}\label{flat}
\mathcal{A}_i\Psi_{0}=0 ,
\end{equation}
in which
\begin{equation}
\mathcal{A}_i\equiv-i\partial_{x_{i}}+i\sum_{j\neq i} \lambda\cot x_{ij} .
\end{equation}

Here and in what follows, unless specified, we shall simply take $L=\pi$ just for convenience.

\section{N=2 Supersymmetric Extension of the Calogero-Sutherland Model}
The N=2 SUSY extended CS model and the $\nu=\frac{5}{2}$ FQHE edge excitations share a common
feature that both can be realized as free boson and fermion CFT in the thermodynamic limit.
As in the previous  section, we write the CS Hamiltonian in the following form
\begin{eqnarray}\label{cs1}
\mathcal{H}&=&\mathcal{A}_i^\dagger \mathcal{A}_i ,\\
\mathcal{A}_i&=&p_i+i\mathbf{G}_i ,\nonumber\\
\mathcal{A}_i^\dagger&=&p_i-i\mathbf{G}_i ,\nonumber\\
p_i&=&-i\partial_{x_i},\nonumber\\
\mathbf{G}_i&=&\sum^N_{j=1,\ j\neq i}\lambda\cot(x_i-x_j),\nonumber\\
\mathbf{G}_{ij} &=& \lambda\cot(x_i-x_j)\nonumber
\end{eqnarray}

There is a N=2 supersymmetric extension to the CS
model\cite{mathieu1}\cite{khare} defined in the
following way
\begin{eqnarray}\label{alpha}
\mathcal{G}&=&\sum^N_{j=1}\theta_j z_j^{-\alpha}( \mathcal{A}_j-\alpha),\nonumber\\
\bar{\mathcal{G}}&=&\sum^N_{j=1}z_j^{-\alpha} (\mathcal{A}_j^\dagger-\alpha)\partial_{\theta_j},\nonumber\\
\mathcal{H}_s&=&\frac{1}{2}\{\mathcal{G},\bar{\mathcal{G}}\}\\
&=&\sum_{i=1}^N-\frac{1}{2}(\partial_i^2+\alpha\partial_i+\partial_i\mathbf{G}_i-\mathbf{G}_i^2)
+\sum_{i,j}[2\alpha(\mathbf{G}_i+\alpha)\delta_{ij}-(\frac{z_j}{z_i})^\alpha\partial_i\mathbf{G}_j]
\theta_i\theta_j^\dag\nonumber
\end{eqnarray}

Here, $\mathcal{G}$ and $\bar{\mathcal{G}}$ are the N=2 SUSY generators, and $\mathcal{H}_s$ is the Hamiltonian.
$\theta_i$ and $\theta_i^\dagger\equiv \partial_{\theta_i}$ are fermion generator and annihilator at
position $i$ respectively.
\begin{equation}\label{theta}
\{\theta_i,\theta_j^\dagger\}=\delta_{i,j} ,\quad \{\theta_i,\theta_j\}=\{\theta_i^\dag,\theta_j^\dag\}=0
\end{equation}

The index $\alpha=0 $ or $\frac{1}{2}$, if the underlying fermionic theory is in the Ramond (R) or Neveu-Schwarz (NS) sector.
We shall elaborate on this point further in our forthcoming paper\cite{yz}. However in the rest of our present paper we shall concentrate ourselves
on the $\alpha=0$ sector.
The vacuum state is annihilated by both $\mathcal{G}$ and $\bar{\mathcal{G}}$ and therefore also by $\mathcal{H}_s$.
Then there are two vacuum states specified by the following conditions respectively,

Case 1)
\begin{eqnarray}\label{vacuum1}
\mathcal{A}_i\tilde{\Psi}_0 &=&  0 ,\nonumber\\
\theta_i^\dagger\tilde{\Psi}_0 &=& 0  ,\nonumber\\
\mathcal{H}_s\tilde{\Psi}_0 &=& 0  ,
\end{eqnarray}
which leads to
\begin{equation}
\tilde{\Psi}_0 = \Psi_0
\end{equation}

Case 2)
\begin{eqnarray}\label{vacuum2}
\mathcal{A}_i^\dagger \tilde{\Psi}_0 &=&  0 ,\nonumber\\
\theta_i \tilde{\Psi}_0&=& 0  ,\nonumber\\
\mathcal{H}_s \tilde{\Psi}_0 &=& 0  ,
\end{eqnarray}
which leads to
\begin{equation}
\tilde{\Psi}_0 =\Psi^{-1}_0\theta_1\theta_2\cdots\theta_N
\end{equation}

Since the ground state in case (1) is more relevant to the FQHE, we
shall consider this case only throughout the present paper. In that
case, similar to the case of CS model, the excited states are
obtained by multiplying the ground state by some polynomials in
$z_i$'s and $\theta_i$'s, which are specifically called Jack
superpolynomials by the authors of ref.\cite{mathieu1}. Here we give a few
examples of the excited eigen-states up to rank 3:

\begin{eqnarray}\label{sjpsi}
  \tilde{\Psi}_{(1,0,0)} &=& b_1\Psi_0 ,\nonumber\\
  \tilde{\Psi}_{(1,1,0)} &=& (b_2-a_1b_1)\Psi_0,\nonumber \\
  \tilde{\Psi}_{(2,0,0)} &=& (b_2+a_1b_1)\Psi_0,\nonumber\\
  \tilde{\Psi}_{(1,1,1)} &=& (b_3-b_2b_1-\frac{1}{2}b_1a_2+\frac{1}{2}b_1a^2_1)\Psi_0 ,\\
  \tilde{\Psi}_{(2,1,0)} &=& (b_2b_1)\Psi_0,\nonumber \\
  \tilde{\Psi}_{(2,1,0)} &=& (b_2a_1-b_1a_2)\Psi_0,\nonumber \\
  \tilde{\Psi}_{(3,0,0)} &=& (b_3+\lambda b_2a_1+\frac{\lambda}{2}b_1a_2+\frac{\lambda^2}{2}b_1a^2_1)\Psi_0,\nonumber
\end{eqnarray}
in which, we have defined
$$a_n=\sum^N_{i=1}z^n_i$$
$$b_n=\sum^N_{i=1}\theta_iz^n_i$$

\section{Generalized Moore-Read State}
In this section we generalize Moore-Read state to the super space formulation.
Moore and Read have written  down the following ground state wave function for $\nu=5/2$ FQHE\cite{read1}.
\begin{equation}
\Psi_{MR}=\mathbf{Pf}(\frac{1}{z_i-z_j})\prod_{i<j}(z_i-z_j)^2 \exp[-\frac{1}{4}\sum|z_i|^2]
\end{equation}

Here, $\mathbf{Pf}(\mathbf{A})$, where $\mathbf{A}$ stands for a $\rm{N}\times N$ antisymmetric matrix, N even, is the Pfaffian of $\mathbf{A}$
defined in the following way,
\begin{equation}
\mathbf{Pf}(\mathbf{A})=
\frac{1}{2^{N/2}[N/2]!}\sum_{\sigma\in S_{N}}{\rm sgn}\,(\sigma)
\mathbf{A}_{\sigma(1),\sigma(2)}\cdots \mathbf{A}_{\sigma(N-1),\sigma(N)}
\end{equation}

which can also be generated in the following pattern

\begin{equation}
(N/2)!(\frac{1}{2}\theta_i \mathbf{A}_{ij} \theta_j)^{N/2}=\mathbf{Pf}(\mathbf{A})\prod_{i=1}^N\theta_i
\end{equation}

Here, $\theta_i$ is the same as defined in eq.(\ref{theta}).

As a corollary, $\mathbf{Pf}^2(\mathbf{A})=\det\mathbf{A}$.

There are two kinds of distinct edge excitations\cite{read2}. one is related to the charge excitation, which
is the same as  Laughlin's quasihole operator, and is symmetric in $z_i$'s.
$$\prod^N_{i=1}(z_i-w)=e_N+\cdots+(-w)^{N-1}e_1+(-w)^N$$
in which
$$e_n=\sum_{1\leq i_1<i_2<\cdots<i_n\leq N}z_{i_1}\cdots z_{i_n}$$

Another set of edge excitations is neutral and antisymmetric in  $z_i$'s. It is related to the breaking
of fermion pairs, as can be seen by the following formula,
\begin{eqnarray}\label{ee}
\lefteqn{\Psi_{n_1,\cdots,n_F}(z_1,\cdots,z_N)=}\nonumber\\
& & \frac{1}{2^{(N-F)/2}[(N-F)/2]!}\sum_{\sigma\in S_{N}}{\rm sgn}\,\sigma
\frac
{\prod_{k=1}^F z_{\sigma(k)}^{n_k}}
{ (z_{\sigma(F+1)}-z_{\sigma(F+2)})\cdots(z_{\sigma(N-1)}-z_{\sigma(N)}) }
\nonumber \\
 & & \qquad\times\prod_{i<j}(z_i-z_j)^qe^{-\frac{1}{4}\sum |z_i|^2}.
\label{ansatzpfedge}
\end{eqnarray}

which can be best expressed  by introducing N anti-commuting $\theta_i$'s as defined in eq.(\ref{theta}):
\begin{eqnarray}\label{theta0}
\lefteqn{\Psi_{n_1,\cdots,n_F}(z_1,\cdots,z_N)\theta_1\theta_2\cdots\theta_N=}  \nonumber\\
&&\frac{1}{2^{(N-F)/2}[(N-F)/2]!}b_{n_1}b_{n_2}\cdots b_{n_F}(\sum_{k\neq l}\theta_k\frac{1}{z_k-z_l}\theta_l)^{\frac{N-F}{2}}
\times\prod_{i<j}(z_i-z_j)^qe^{-\frac{1}{4}\sum |z_i|^2}.
\end{eqnarray}

From the discussions in the proceeding section, we see that the
ground state for N=2 SUSY extended CS model does not contain any
Pfaffian factor, therefore it does not constitute the ground state
of the $\nu=5/2$ edge excitations. At first sight, it seems hopeless
to relate such a theory to the $\nu=5/2$ edge excitations in FQHE.
On the other hand, the introduction of the fermion creators and
annihilators in the equation above, eq.(\ref{theta0}), seems on the
right track in dealing with antisymmetric N-variable functions, just
like what we have seen in BCS superconductivity theory.

As a first step, we may just introduce N fermionic variable
$\theta_i$'s and their hermitian conjugates $\theta_i^\dagger$'s by
extending the ground state wave function in CS model to the
superspace formalism.

\begin{eqnarray}\label{ssin}
\tilde{\Psi}&=&\exp\{\lambda\sum_{i<j}{\log (\sin X_{ij})}\},\nonumber\\
&=&\Psi_0  \exp\{-\frac{1}{2}\theta \mathbf{G}\theta\}.
\end{eqnarray}
Here, $\Psi_0$ , the bosonic part of $\tilde{\Psi}$, is the same as defined in eq.(\ref{psi})
\begin{eqnarray*}
X_{ij}=x_{ij}-\theta_i\theta_j
\end{eqnarray*}
is called the superdistance in superspace.

Notice that the usual superspace formalism is only valid in dealing
with relativistic SUSY, with Hamiltonian linear in dispersion
relation. And in that case we can also define covariant super
derivative in superspace.

The Pfaffian like state,
eq.(\ref{ssin}), is not N=2 nonrelativistic SUSY invariant ground state
because $\theta_i^\dagger$ does not annihilate it. Nevertheless, if we
expand eq.(\ref{ssin}) in $\theta_i$ variable, we do find P-wave
pairing similar in BCS theory and in edge excitations of the
Moore-Read state. The only difference from the later is that the
number of the fermion pairs is not fixed.

\begin{eqnarray}
 \exp\{-\frac{1}{2}\theta \mathbf{G}\theta\}
&=&\prod^N_{i=1,i<j}(1-\theta_i \mathbf{G}_{ij}\theta_j) ,\nonumber\\
&=&\sum^N_{n=0}\sum_{\{m_1<\cdots <m_{2n}\}}(-1)^n\theta_{m_1}\cdots
\theta_{m_{2n}}\mathbf{Pf}(\mathbf{G}_{\{m_1\cdots m_{2n}\}}).
\end{eqnarray}
Here, $\mathbf{G}_{\{m_1\cdots m_{2n}\}},n \leq N/2$, is an $2n \times 2n$
antisymmetric matrix with matrix elements
\begin{equation}
(\mathbf{G}_{\{m_1\dots m_{2n}\}})_{ij}=\mathbf{G}_{m_i,m_j}
\end{equation}

\section{Bogoliubov Transformation and the New Vacuum State}

This section contains our main formalism for solving edge
excitations with respect to Moore-Read state. We have found that
just like the case of odd denominator FQHE, the edge excitations for
$\nu=5/2$ is also factorized as the product of the ground state wave
function and some polynomial functions.

First, notice that if we treat eq.(\ref{ssin}) as the definition of the ground state wave function, then because of the fermion pairing, this wave function is not annihilated by the
$\theta_i^\dagger$'s.
The situation looks similar in what happens in BCS theory. So we make the following Bogoliubov transformation such that the wave function as defined in eq.(\ref{ssin}) is
annihilated by the new fermion annihilators $\tilde{\theta}_i^{\dag}$,

\begin{equation}\label{bogo}
\begin{array}{lll}
\tilde{x}_i &=& x_i \\
\tilde{\theta}_i  &=& \left( \frac{1}{1-\mathbf{G}}\right)_{ij}   (\theta_j + \mathbf{G}_{jk}\theta_k^{\dag})\\
\tilde{\theta}_i^{\dag}  &=& \left(  \frac{1}{1-\mathbf{G}}
\right)_{ij} (\theta_j^{\dag} + \mathbf{G}_{jk}\theta_k) .
\end{array}.
\end{equation}

It can be checked that $\det(1+\mathbf{G})\geq 1$, so that the coordinate transformation eq.(\ref{bogo}) is
regular every where in coordinate space $\{x_i\}$.
Eq.(\ref{bogo}) is called the Bogoliubov transformation because it can be checked that, indeed, the following conditions are satisfied,
\begin{equation}\label{comp}
\begin{array}{lll}
\{\tilde{\theta}_i^{\dag} , \tilde{\theta}_j \} &=& \delta_{ij} ,\quad \{\tilde{\theta}_i , \tilde{\theta}_j \}=
\{\tilde{\theta}_i^{\dag} , \tilde{\theta}_j^\dag \}=0\\
\left[ \partial_{\tilde{x}_i},\tilde{\theta}_j\right]  &=& \left[ \partial_{\tilde{x}_i},\tilde{\theta}_j^{\dag} \right] =0\\
\end{array}
\end{equation}

For detailed proof of eq.(\ref{comp}), see appendix A.

The new vacuum state with respect to the new set of fermion
oscillators $\tilde{\theta}_i$ and $\tilde{\theta}_i^{\dag}$ is
defined by
\begin{equation}\label{nv}
\tilde{\theta}_i^{\dag} |\tilde{0}\rangle=0  ,\quad
\langle\tilde{0}|\tilde{0}\rangle= 1 .
\end{equation}

Eq.(\ref{nv}) leads to the following solution
\begin{equation}\label{vac}
|\tilde{0}\rangle=\det(1+\mathbf{G})^{-\frac{1}{2}}
e^{-\frac{1}{2}\theta \mathbf{G}\theta} |0\rangle
\end{equation}
Here $|0\rangle $ is the vacuum state with respect to $\theta_i$ and $\theta_i^\dagger$,
\begin{equation}
\det(1+\mathbf{G})^{-\frac{1}{2}}=\langle0|e^{\frac{1}{2}\theta^{\dag} \mathbf{G}\theta^{\dag}}e^{-\frac{1}{2}\theta \mathbf{G}\theta}|0\rangle^{-\frac{1}{2}}
\end{equation}
is the normalization factor as can be checked as following

\begin{eqnarray}\label{norm}
&&\langle0|e^{\frac{1}{2}\theta^{\dag} \mathbf{G}\theta^{\dag}}e^{-\frac{1}{2}\theta \mathbf{G}\theta}|0\rangle \nonumber\\
&=&\sum^N_{n=0}\sum_{\{m_1<\cdots <m_{2n}\}}(-1)^n\mathbf{Pf}^2(\mathbf{G}_{\{m_1\cdots m_{2n}\}})\langle0|\theta^{\dag}_{m_1}\cdots \theta^{\dag}_{m_{2n}}\theta_{m_1}\cdots \theta_{m_{2n}}|0\rangle \nonumber\\
&=&\sum^N_{n=0}\sum_{\{m_1<\cdots <m_{2n}\}}\mathbf{Pf}^2(\mathbf{G}_{\{m_1\cdots m_{2n}\}})\nonumber\\
&=&\sum^N_{n=0}\sum_{\{m_1<\cdots <m_{2n}\}}\det(\mathbf{G}_{\{m_1\cdots m_{2n}\}})\nonumber\\
&=&\det(1+\mathbf{G}).\\
&&E.O.F.\nonumber
\end{eqnarray}
Notice that the normalized fermion vacuum  $|\tilde{0}\rangle $, is invariant under the $SO(N)$ transformation
\begin{eqnarray}\label{va1}
|\tilde{0}\rangle &=&e^{-\frac{1}{2} \theta \mathbf{G} \theta} \det(1+\mathbf{G})^{-\frac{1}{2}}|0\rangle\nonumber\\
&=&e^{-\frac{1}{2} \bar{\theta} \mathbf{G} \bar{\theta}} \det(1+\mathbf{G})^{-\frac{1}{2}}|0\rangle ,
\end{eqnarray}
where $\bar{\theta}=\frac{1+\mathbf{G}}{1-\mathbf{G}}\theta$, $\frac{1+\mathbf{G}}{1-\mathbf{G}}\in SO(N)$
as can be easily checked.

Combining the Jastrow factor in the charged sector with the new fermion vacuum, we finally arrive at the following generalized and "normalized" Moore-Read state on the edge,
\begin{equation}
\tilde{\Psi}_0=\Psi_0  \exp\{-\frac{1}{2}\theta \mathbf{G}\theta\}\det(1+\mathbf{G})^{-\frac{1}{2}}
\end{equation}
Here ``normalized'' only means the new fermion vacuum is normalized to unity.

Now it is straightforward to write down the N=2 supersymmetric Hamiltonian
in terms of the new fermionic variables $\tilde{\theta}_i$'s, $\tilde{\theta}_i^{\dag}$'s, and the
new momentum operator $\tilde{p}_i$'s, just as eq.(\ref{alpha}),

\begin{eqnarray}
\tilde{\mathcal{G}} &=&\sum_{j=1}^N -i\tilde{\theta}_i(\partial_{\tilde{x}_i}- \mathbf{G}_i),\nonumber\\
\tilde{\bar{\mathcal{G}}}&=&\sum_{j=1}^N -i(\partial_{\tilde{x}_i}+\mathbf{G}_i)\tilde{\theta}_i^{\dag},\nonumber\\
\tilde{\mathcal{H}}&=&\frac{1}{2}\{\tilde{\mathcal{G}}, \tilde{\bar{\mathcal{G}}} \}\\
&=&-\frac{1}{2}\sum_{i=1}^N\partial_{\tilde{x}_i}^2+\sum_{i<j}\frac{\lambda(\lambda-1
+\tilde{\theta}_{ij}\tilde{\theta}_{ij}^\dag)}{\sin^2x_{ij}}-E_0 .\nonumber
\end{eqnarray}

It is trivial to check that the following conditions are satisfied
\begin{equation}
\tilde{\mathcal{G}}^2=\tilde{\bar{\mathcal{G}}}^2=\tilde{\mathcal{G}}\tilde{\Psi}_0=\tilde{\bar{\mathcal{G}}}\tilde{\Psi}_0=0
\end{equation}
as required by the N=2 non-relativistic SUSY discussed in section
III, provided the following comparability conditions  are satisfied,
\begin{equation}
[\partial_{\tilde{x}_i},\tilde{\theta}_j^{\dag}]=\partial_{\tilde{x}_i}|\tilde{0}\rangle=0\\
\end{equation}

We leave the detailed proof of the compatibility check in Appendix
A. Here we just give the final result for the transformation of the
momentum operator $p_i$'s as derived from eq.(\ref{bogo}).
\begin{equation}\label{partial}
\partial_{\tilde{x}_i} =\partial_{x_i} +\frac{1}{2}
(\theta+\theta^{\dag})\mathbf{B}^T \mathbf{G}_{,i} \mathbf{B}(\theta+\theta^{\dag})
\end{equation}
Here, $\mathbf{G}_{,i}$ means the $x_i$ derivative of the matrix $\mathbf{G}$ and
\begin{equation}
\begin{array}{lll}
 \mathbf{B}\equiv \frac{1}{1-\mathbf{G}} &, & \mathbf{B}^T  \equiv \frac{1}{1+\mathbf{G}}
\end{array}
  \end{equation}

Just for completeness, we write down the N=2 Hamiltonian in terms of the old variables,
\begin{eqnarray}\label{2s}
 \mathcal{\tilde{G}}&=&\sum_{j=1}^N(\mathbf{B\theta+BG\theta^\dag})_j[-i \partial_{x_j} -\frac{i}{2}  (\theta+\theta^{\dag}) \mathbf{B}^T \mathbf{G}_{,j} \mathbf{B}(\theta+\theta^{\dag})+i\mathbf{G}_j], \nonumber\\
 \tilde{\bar{\mathcal{G}}}&=&\sum_{j=1}^N[-i \partial_{x_j} -\frac{i}{2}  (\theta+\theta^{\dag}) \mathbf{B}^T \mathbf{G}_{,j} \mathbf{B}(\theta+\theta^{\dag})-i\mathbf{G}_j](\mathbf{B\theta^\dag+BG\theta})_j ,\nonumber\\
\mathcal{\tilde{H}}&=&\frac{1}{2}\{\tilde{\mathcal{G}},\tilde{\bar{\mathcal{G}}}\}\\
&=&-\frac{1}{2}\sum_{i=1}^N\{\partial_i^2+(\theta+\theta^{\dag}) \mathbf{B}^T \mathbf{G}_{,i} \mathbf{B}(\theta+\theta^{\dag})\partial_{x_i}
+\frac{1}{4}[(\theta+\theta^{\dag}) \mathbf{B}^T \mathbf{G}_{,i} \mathbf{B}(\theta+\theta^{\dag})]^2\nonumber\\
&&+\frac{1}{2}(\theta+\theta^{\dag})[\mathbf{B}^T \mathbf{G}_{,i}\mathbf{B}]_{,i}(\theta+\theta^{\dag})\}-E_0\nonumber\\
&&+\sum_{i<j}\frac{1}{\sin^2x_{ij}}\{\lambda(\lambda-1)
+2(\mathbf{B\theta+BG\theta^\dag})_i[(\mathbf{B\theta^\dag+BG\theta})_i-(\mathbf{B\theta^\dag+BG\theta})_j]\}\nonumber
\end{eqnarray}

And the ground state wave function
\begin{eqnarray}\label{tpsi}
\tilde{\Psi}_0 &= &\psi_0 e^{-\frac{1}{2} \theta \mathbf{G} \theta} e^{-\frac{1}{2} \mathrm{Tr}\ln (1+\mathbf{G})},\nonumber\\
&=& \prod_{i<j}\sin^{\lambda}[(x_{i}-x_{j})]e^{-\frac{1}{2} \theta \mathbf{G} \theta} \det(1+\mathbf{G})^{-\frac{1}{2}}.
\end{eqnarray}

The supersymmetric generators $\mathcal{\tilde{G}}$ and $\tilde{\bar{\mathcal{G}}}$ as well as the
Hamiltonian $\mathcal{\tilde{H}}$, eq.(\ref{2s}), appears complicated in electronic coordinates
$x_i$,  $p_i$, $\theta_i$, $\theta_i^\dag$. One may ask a natural question if the system is integrable.
Our answer is definitely positive. The reason is that the system can be written as the standard form of the N=2
non-relativistic SUSY system, the integrability of the later is proven in ref.\cite{mathieu1} and refs therein.
Thus the integrability of the system specified by eq.(\ref{2s}) is ensured, provided that the coordinate
transformation eq.(\ref{bogo}) and eq.(\ref{partial}) is smooth and nonsigular everywhere. The regularity of
eq.(\ref{bogo}) and eq.(\ref{partial}) is guaranteed since for real antisymmetric matrix $\mathbf{G}$,
\begin{equation}
\det(1-\mathbf{G})=\det(1+\mathbf{G})\geq 1,
\end{equation}
as can be seen from eq.(\ref{norm}).

The excited states in our N=2 SUSY CS
model, are obtained by acting on the $\tilde{\Psi}_0$, with
the $a_n$'s and $\tilde{b}_n$'s. The first a few low lying eigenstates would appear in the 
same form as specified by eq.(\ref{sjpsi}), with $b_m$'s replaced by
$\tilde{b}_m$'s and $\Psi_0$ by $\tilde{\Psi}_0$. 
The edge excitations as specified in eq.(\ref{ee}) or
eq.(\ref{theta0}), however, become  products of the bosonic and fermionic
oscillators, $a_n$'s and $b_m$'s, with the ground state wave
function  $\tilde{\Psi}_0 $. 
It remains unclear if the two procedures produce
the same results. But after examining the
fewer number of $\tilde{b}_n$ excitations, we found that, at least
for the coefficient of the expanded term $\prod_{i=1}^N \theta_i$,
the corresponding edge excitations contains either some symmetric
function multiplied by the Pfaffian state, or a sum of products of
the antisymmetric functions with the sub-Pfaffian states. By
sub-Pfaffian , we mean the Pfaffians of the lower dimensional
antisymmetric matrices. In conclusion, the edge excitations in our
formalism are generalizations to the Moore-Read's form.

To find the relation between our edge excitations  and the Moore and Read's, here we work out explicitly an example which contains two "quasi hole" excitations:
\begin{eqnarray}\label{ee0}
&&\tilde{b}_{n_1}\tilde{b}_{n_2} |\tilde{0}\rangle    ,\nonumber\\
&=& [\mathbf{z}^{n_1}(\mathbf{B}\theta+\mathbf{BG}\theta^{\dag})][\mathbf{z}^{n_2}(\mathbf{B}\theta+\mathbf{BG}\theta^{\dag})]  |\tilde{0}\rangle ,\nonumber\\
&=&\{\mathbf{z}^{n_1}[(\mathbf{B(1-G^2)}\theta+\mathbf{G}\tilde{\theta}^{\dag}]\}\{\mathbf{z}^{n_1}[(\mathbf{B(1-G^2)}\theta+\mathbf{G}\tilde{\theta}^{\dag}]\}  |\tilde{0}\rangle ,\nonumber\\
&=&(B_{n_1}+\mathbf{z^{n_1}G}\tilde{\theta}^\dag)(B_{n_2}+\mathbf{z^{n_2}G}\tilde{\theta}^\dag) |\tilde{0}\rangle\nonumber\\
&=&(:\tilde{b}_{n_1}\tilde{b}_{n_2}:+F_{n_1n_2}) |\tilde{0}\rangle .
\end{eqnarray}
Here, $B_{n_1}\equiv\mathbf{z}^{n_1}\mathbf{(1+G)}\theta$, $F_{n_1n_2}\equiv \{\mathbf{z^{n_1}G}\tilde{\theta}^\dag,B_{n_2}\}
=\mathbf{z^{n_1}Gz^{n_2}}$, and
\begin{eqnarray}\label{va}
|\tilde{0}\rangle &=&e^{-\frac{1}{2} \theta \mathbf{G} \theta} \det(1+\mathbf{G})^{-\frac{1}{2}}|0\rangle
\end{eqnarray}
is the normalized fermion vacuum with respect to $\tilde{\theta}_i$
and $\tilde{\theta}_i^\dag$, and  $|0\rangle$ w.r.t. $\theta_i$ and
$\theta_i^\dag$.
A ``normal ordering'' is defined with respect to $|\tilde{0}\rangle$,
\begin{equation}\label{normal}
:\tilde{\theta_i}\tilde{\theta}_j^\dag:=-:\tilde{\theta}_j^\dag\tilde{\theta_i}:=\tilde{\theta_i}\tilde{\theta}_j^\dag
\end{equation}
The term $F_{n_1n_2}$ can be considered as the Wick contraction term. By the definition of eq.(\ref{normal}), we have
\begin{equation}
:\tilde{b}_{n_1}\tilde{b}_{n_2}\cdots\tilde{b}_{n_{F-1}}\tilde{b}_{n_F}: |\tilde{0}\rangle=
B_{n_1} \cdots B_{n_F} |\tilde{0}\rangle ,
\end{equation}
where, use has been made of the fact that $\tilde{\theta}_i^\dag |\tilde{0}\rangle=0$.
Applying Wick's contraction rules, we have, for the $F$ quasi hole excitations,

\begin{eqnarray}\label{ee1}
&&\tilde{b}_{n_1}\tilde{b}_{n_2}\cdots\tilde{b}_{n_{F-1}}\tilde{b}_{n_F} |\tilde{0}\rangle   \nonumber\\
&=&\sum_{l\atop(F-l)\, even}\sum_{\sigma\in S_l}
{\rm sgn}\,\sigma\frac{1}{l!2^{\frac{F-l}{2}}(\frac{F-l}{2})!}B_{n_{\sigma(1)}}\cdots B_{n_{\sigma(l)}}
F_{n_{\sigma(l+1)}n_{\sigma(l+2)}}\cdots F_{n_{\sigma(F-1)}n_{\sigma(F)}} |\tilde{0}\rangle .
\end{eqnarray}

The term $F_{nm}$ on the r.h.s. of eq.(\ref{ee0}), can be worked out, assuming $n>m$  without losing of generality,
\begin{eqnarray*}
&&\mathbf{z}^m\mathbf{G}\mathbf{z}^n\\
&=& iz^m_i\frac{z_i+z_j}{z_i-z_j}z^n_j\\
&=& -i(n-m)a_{m+n}+i\sum^{n-m-1}_{l=0}a_{n-l}a_{m+l}.
\end{eqnarray*}

So the term $\mathbf{z}^m\mathbf{G}\mathbf{z}^n$ represents bosonic excitations in terms of $a_i$'s, which are in the charged sector,
provided we take $\theta_i$'s and $\theta_i^\dag$'s as the canonical fermion variables.

To compare our results with More and Read's, we can just consider the term
\begin{equation}\label{ee3}
\tilde{b}_{n_1}\tilde{b}_{n_2}\cdots\tilde{b}_{n_{F-1}}\tilde{b}_{n_F} |\tilde{0}\rangle = \tilde{b}_{n_1}\tilde{b}_{n_2}\cdots\tilde{b}_{n_{F-1}}\tilde{b}_{n_F}  e^{-\frac{1}{2} \theta \mathbf{G} \theta} \det(1+\mathbf{G})^{-\frac{1}{2}}|0\rangle ,
\end{equation}
and expand the r.h.s. of eq.(\ref{ee3}) in $\theta_i$ powers. From the term which contains the maximal number of $\theta_i$'s ,
we get
\begin{eqnarray}\label{theta1}
&&\tilde{b}_{n_1}\tilde{b}_{n_2}\cdots\tilde{b}_{n_{F-1}}\tilde{b}_{n_F}
e^{-(\frac{1}{2}\theta \mathbf{G} \theta)}|_{\mbox{terms with maximal number of }\theta_i's}\nonumber\\
&=&\sum_{l=0}^F\sum_{\sigma\in S_l}{\rm sgn}\,\sigma
\frac{1}{l!2^{\frac{F-l}{2}}(\frac{F-l}{2})!}(-1)^{\frac{N-l}{2}}f_{i_1}^{n_{\sigma(1)}}\cdots
f_{i_l}^{n_{\sigma(l)}}  \theta_{i_1}\cdots \theta_{i_l}
F_{n_{\sigma(l+1)}n_{\sigma(l+2)}} \cdots F_{n_{\sigma(F-1)}n_{\sigma(F)}}\nonumber\\
&&\times \theta_{i_{l+1}}\cdots \theta_{i_N}G_{l+1 l+2} \cdots G_{N-1 N}\nonumber\\
&=&\sum_{l=0}^F\sum_{\rho\in S_N}{\rm sgn}\,\rho \sum_{\sigma\in S_l}{\rm sgn}\,\sigma
\frac{1}{l!2^{\frac{F-l}{2}}(\frac{F-l}{2})!}(-1)^{\frac{N-l}{2}}\theta_1\cdots\theta_N
f_{\rho(1)}^{n_\sigma(1)} \cdots f_{\rho(l)}^{n_\sigma(l)}\nonumber\\
&&\times G_{\rho(l+1) \rho(l+2)} \cdots G_{\rho(N-1) \rho(N)} F_{n_{\sigma(l+1)}n_{\sigma(l+2)}} \cdots F_{n_{\sigma(F-1)}n_{\sigma(F)}},
\end{eqnarray}
where
\begin{equation}
f_i^n=z_j^n(1+\mathbf{G})_{ji}
\end{equation}

Besides the normalization factor,
$\det(1+\mathbf{G})^{-\frac{1}{2}}$, the form of eq.(\ref{theta1}),
looks differently from Moore-Read edge excitations taking place
on the unit circle. The rather complicated form on the r.h.s. of
eq.(\ref{theta1}) may indicate that the charge and the spin excitations
are correlated in fact.

\section{Conclusions and Speculations}
In summary, we have generalized Moore-Read state and its edge
excitations in $\nu=5/2$ FQHE to the superspace formalism, by
introducing a set of fermion oscillators. We have found that the
Hamiltonian for the corresponding edge excitations is N=2
supersymmetric. In terms of Bogoliubov transformed fermion and bose
variables, the Hamiltonian is simply the one for the N=2 extended
Calogero-Sutherland model. In this paper, we have shown how SUSY in
both relativistic and non-relativistic theory are in fact
correlated. In the first sight, they  look quite different. In the
formal formalism, $\theta$ variable as a Grassmann number, is
regarded as a fermionic coordinate in superspace. And its role is
purely auxiliary, in the sense that we could formulate SUSY
transformations  completely in component fields, forgetting about
$\theta$. While in the later case, it acts as fermion creator and
annihilator, and enters Hamiltonian as an indispensable part.
However, the relation become more transparent if we make the
following observation. From the point of view of relativistic
conformal field theory, we can regard $\theta_i$ as an external
fermionic source coupled to fermion field in one spatial dimension,
then $\theta_i$  can be considered as the boundary value of the
fermion field in two spatial dimensions. In the thermodynamic limit,
the bulk low energy effective theory  is dual to a relativistic CFT
on the boundary, if holographic principle works here. We shall
elaborate on this point further on our forthcoming paper\cite{yz}.
The relation between relativistic and non-relativistic SUSY theory
is made even more explicit by using Bogoliubov transformation, which
transform the correlators in one spatial  dimensional superconformal
field theory into the vacuum state in non-relativistic N=2 SUSY
theory. This is the main point in our present paper.

\begin{acknowledgments}
It is pleasant to acknowledge the useful discussions with Lu Yu and Yue Yu.
Special thanks go to Yue Yu, whose lecture on the fractional quantum
Hall effects has played an important role in drawing our attention
into this field. Discussions with Fanrong Xu at the preliminary
stage of the present work is also acknowledged. This work is
supported in part by grants from the Chinese Academy of Sciences.
\end{acknowledgments}

\section{Appendix A}

In what follows we shall check explicitly that the compatibility
conditions, eq.(\ref{comp}), are indeed satisfied. To simplify our
proof, let us first define the following matrix
\begin{equation}
\begin{array}{lll}
 \mathbf{B}\equiv \frac{1}{1-\mathbf{G}} &, & \mathbf{B}^T  \equiv \frac{1}{1+\mathbf{G}}
\end{array}
  \end{equation}

Then the Bogoliubov transformation, eq.(\ref{bogo}), and its inverse can be simply written as
\begin{eqnarray}
 \tilde{\theta} &=& \mathbf{B}(\theta+\theta^{\dag})-\theta^{\dag}\\
\tilde{\theta}^{\dag} &=& \mathbf{B}(\theta+\theta^{\dag})-\theta
\end{eqnarray}
and
\begin{eqnarray}
 \theta &=& \mathbf{B}^T (\tilde{\theta}+\tilde{\theta}^{\dag})-\tilde{\theta}^{\dag}\\
\theta^{\dag} &=& \mathbf{B}^T
(\tilde{\theta}+\tilde{\theta}^{\dag})-\tilde{\theta}
\end{eqnarray}

We can then work out the transformation for $p_i$'s
\begin{eqnarray}\label{px}
\partial_{\tilde{x}_i} &=& \frac{\partial x_j}{\partial \tilde{x}_i} \partial x_j
+ \frac{\partial \theta_j}{\partial \tilde{x}_i} \partial \theta_j +
\frac{\partial \theta^{\dag}_j}{\partial \tilde{x}_i} \partial\theta_j^{\dag} ,\nonumber\\
&=&\partial_{x_i} +[\mathbf{B}^T_{,i}(\tilde{\theta}+\tilde{\theta}^{\dag})]_j (\partial_{\theta_j} +\partial_{\theta_j^{\dag}}),\nonumber\\
&=&\partial_{x_i} +[\mathbf{B}^T_{,i}(2\mathbf{B}-1) (\theta+\theta^{\dag})]_j (\partial_{\theta_j} +\partial_{\theta_j^{\dag}}),\nonumber\\
&=&\partial_{x_i}-(\mathbf{B}^T\mathbf{G}_{,i}\mathbf{B})_{jk}(\theta+\theta^{\dag})_k (\partial_{\theta_j} +\partial_{\theta_j^{\dag}}).
\end{eqnarray}
Notice that $\mathbf{B}^T\mathbf{G}_{,i}\mathbf{B}$ is an antisymmetric matrix.
Now we may make the following identification
 \begin{equation}
\begin{array}{lll}
 \partial_{\theta_i} & \longrightarrow& \theta_i^{\dag}\\
\partial_{\theta_i^{\dag}} & \longrightarrow& \theta_i
\end{array} .
\end{equation}
But if we make the above identification, then $\theta_j+\theta^{\dag}_j$ in eq.(\ref{px}) could also be identified as
$\partial_{\theta_j}+\partial_{\theta_j^\dagger}$. To avoid double counting of the derivatives, we could just multiply the second term in the last line of eq.(\ref{px})
by a factor of $\frac{1}{2}$, while changing $\partial_{\theta_j} +\partial_{\theta_j^{\dag}}$ to $\theta+\theta^{\dag}$.

Hence we arrive at the following transformation derived  from the Bogoliubov transformation, eq.(\ref{bogo}),
\begin{equation}
\partial_{\tilde{x}_i} =\partial_{x_i} +\frac{1}{2}
(\theta+\theta^{\dag})\mathbf{B}^T \mathbf{G}_{,i} \mathbf{B}(\theta+\theta^{\dag})
\end{equation}

The correctness of the argument we have just made above can be
verified by checking the following compatibility conditions. First
we check,
\begin{equation}
\begin{array}{lll}
 &&\left[ \partial_{\tilde{x}_i}, \tilde{\theta}_l\right] \\
&=&[\partial_{x_i} +\frac{1}{2}(\theta+\theta^{\dag}) \mathbf{B}^T \mathbf{G}_{,i} \mathbf{B}(\theta+\theta^{\dag}),(\mathbf{B}(\theta+\theta^{\dag})-\theta^{\dag})_l]\\
&=&[\mathbf{B}_{,i}(\theta+\theta^{\dag})]_l
+\frac{1}{2}[(\theta+\theta^{\dag})_j\mathbf{C}_{jk}(\theta+\theta^{\dag})_k
, \mathbf{B}_{lm}
(\theta+\theta^{\dag})_m-\theta^{\dag}_l]\\
&=&[\mathbf{B}_{,i}(\theta+\theta^{\dag})]_l
+\frac{1}{2}(\theta+\theta^{\dag})_j \mathbf{C}_{jk}
(2\mathbf{B}_{lk}-\delta_{lk})
-\frac{1}{2}(2\mathbf{B}_{lj}-\delta_{lj})\mathbf{C}_{jk}(\theta+\theta^{\dag})_k\\
&=&[\mathbf{B}_{,i}(\theta+\theta^{\dag})]_l+\frac{1}{2}[(\theta+\theta^{\dag})(\mathbf{B}^T\mathbf{G}_{,i}\mathbf{B})(2\mathbf{B}^T-1)]_l
-\frac{1}{2}[(2\mathbf{B}-1) (\mathbf{B}^T\mathbf{G}_{,i}\mathbf{B})(\theta+\theta^{\dag})]_l\\
&=&[\mathbf{B}_{,i}(\theta+\theta^{\dag})]_l+\frac{1}{2}[(\theta+\theta^{\dag})\mathbf{B}^T\mathbf{G}_{,i}\mathbf{B}^T]_l-\frac{1}{2}[\mathbf{B}\mathbf{G}_{,i}\mathbf{B}(\theta+\theta^{\dag})]_l\\
&=&[\mathbf{B}_{,i}(\theta+\theta^{\dag})]_l-[\mathbf{B}\mathbf{G}_{,i}\mathbf{B}(\theta+\theta^{\dag})]_l\\
&=&0
\end{array}
\end{equation}
as desired.

Similarly we have:  $[\partial_{\tilde{x}_i},\tilde{\theta}_j^{\dag}]=0$\\
Second, we shall also check if $\partial\tilde{x}_i$ kills the new fermion vacuum. We have
\begin{equation}\label{px0}
\partial_{\tilde{x}_i} |\tilde{0}\rangle
=[\partial_{x_i} +\frac{1}{2}
(\theta+\theta^{\dag})\mathbf{B}^T \mathbf{G}_{,i} \mathbf{B}(\theta+\theta^{\dag})]\det(1+\mathbf{G})^{-\frac{1}{2}}
e^{-\frac{1}{2}\theta \mathbf{G}\theta} |0\rangle
\end{equation}

Let us first consider the second term in the first bracket on the r.h.s. of eq.(\ref{px0})
acting on the fermion vacuum.
To simplify the notation, we shall define
\begin{equation}
\mathbf{C}\equiv \mathbf{B}^T\mathbf{G}_{,i}\mathbf{B}
\end{equation}
Now we proceed
\begin{eqnarray}\label{px2}
&& \frac{1}{2}[(\theta+\theta^\dag)\mathbf{B}^T \mathbf{G}_{,i} \mathbf{B}(\theta+\theta^{\dag})] |\tilde{0}\rangle\nonumber\\
&=& [\frac{1}{2}\theta_j \mathbf{C}_{jk}(\theta_k+  \theta_m \mathbf{G}_{mk})+\frac{1}{2}  \mathbf{C}_{jk} \mathbf{G}_{jk}
+\frac{1}{2}\mathbf{C}_{jk} (\theta_k + \theta_m \mathbf{G}_{mk}) \theta_l \mathbf{G}_{lj} ] |\tilde{0}\rangle\nonumber\\
&=&\frac{1}{2} [\theta \mathbf{C} \theta-  \theta \mathbf{C} \mathbf{G} \theta +
 \theta \mathbf{G} \mathbf{C} \theta -  \mathrm{Tr}(\mathbf{C}\mathbf{G})-
 \theta \mathbf{G}\mathbf{C}\mathbf{G} \theta ] |\tilde{0}\rangle\nonumber\\
&=& \frac{1}{2}[\theta\frac{1}{1+\mathbf{G}} \mathbf{G}_{,i} \theta
- \mathrm{Tr}(\mathbf{C}\mathbf{G}) +\theta \mathbf{G} \frac{1}{1+\mathbf{G}} \mathbf{G}_{,i}\theta]
|\tilde{0}\rangle\nonumber\\
&=& \frac{1}{2} [\theta \mathbf{G}_{,i} \theta -  \mathrm{Tr}(\mathbf{C}\mathbf{G})]
|\tilde{0}\rangle\nonumber\\
&=&  [\frac{1}{2} \theta \mathbf{G}_{,i} \theta +\frac{1}{2} \partial_i \mathrm{Tr} \ln(1+\mathbf{G})]|\tilde{0}\rangle
\end{eqnarray}

And the first term in the first bracket on the r.h.s. of eq.(\ref{px0}) acting on the new fermion vacuum,
gives us just the result,
\begin{equation}\label{px1}
\partial_{x_i} [\det\left(1+\mathbf{G}\right)^{-\frac{1}{2}}
e^{-\frac{1}{2}\theta \mathbf{G}\theta}] |0\rangle=
[- \frac{1}{2} \theta \mathbf{G}_{,i} \theta -\frac{1}{2} \partial_i \mathrm{Tr} \ln\left(1+\mathbf{G}\right)]
\det\left(1+\mathbf{G}\right)^{-\frac{1}{2}}e^{-\frac{1}{2}\theta \mathbf{G}\theta}|0\rangle .
\end{equation}

Combining eq.(\ref{px2}) and eq.(\ref{px1}), we then have
\begin{equation}
\partial_{\tilde{x}_i} |\tilde{0}\rangle=0 ,
\end{equation}
as the desired result.

Finally we make a check if the  $\tilde{\theta_i}$'s and
$\tilde{\theta_i^\dag}$'s are in an orthonormal set, such that they are
equally good fermion variables just as the old ones.
\begin{eqnarray*}
 &&\{\tilde{\theta}_i^{\dag} , \tilde{\theta}_k\}\\
& =& \left(\frac{1}{1-\mathbf{G}}\right)_{ij} \left(\frac{1}{1-\mathbf{G}}\right)_{kl}
\{\left(\theta^{\dag} + \mathbf{G} \theta\right)_j , \left(\theta+\mathbf{G} \theta^{\dag}\right)_l\}\\
&=& \left(\frac{1}{1-\mathbf{G}}\right)_{ij} \left(\frac{1}{1-\mathbf{G}}\right)_{kl} \left(1 - \mathbf{G}^2\right)_{jl}\\
&=& \left(1+\mathbf{G}\right)_{il} \left(\frac{1}{1+\mathbf{G}}_{lk}\right)\\
&=& \delta_{ik}
\end{eqnarray*}
And
\begin{eqnarray*}
 &&\{\tilde{\theta}_i , \tilde{\theta}_k\}\\
& =& \left(\frac{1}{1-\mathbf{G}}\right)_{ij} \left(\frac{1}{1-\mathbf{G}}\right)_{kl}
\{\left(\theta + \mathbf{G} \theta^{\dag} \right)_j , \left(\theta+\mathbf{G} \theta^{\dag}\right)_l\}\\
&=& \left(\frac{1}{1-\mathbf{G}}\right)_{ij} \left(\frac{1}{1-\mathbf{G}}\right)_{kl} \left(\mathbf{G}_{lj}+\mathbf{G}_{jl}\right)\\
&=& 0
\end{eqnarray*}

Similarly, we have
$$\{\tilde{\theta}_i^\dag , \tilde{\theta}_k^\dag\}=0 .$$
Hence the SUSY extended Moore-Read state on the edge, as defined in eq.(\ref{tpsi}), satisfies the following conditions,
\begin{eqnarray}
[-i \partial_{x_i} -\frac{i}{2}  \left(\theta+\theta^{\dag}\right) \mathbf{B}^T \mathbf{G}_{,i} \mathbf{B}\left(\theta+\theta^{\dag}\right)-i\mathbf{G}_i ] \tilde{\Psi}_0 &=& 0\\
\frac{1}{1-\mathbf{G}} \left(\theta^{\dag} + \mathbf{G} \theta\right)
\tilde{\Psi}_0 &=& 0
\end{eqnarray}

\section{Note Added}
After the completion of the present work, the following ref.\cite{hasebe2},
was brought to our attention. The framework of \cite{hasebe2}
is quite different from ours. Nevertheless, we found some overlaps in the
final results.
\section{Note Added in Revised Edition}
After the submission of the paper in previous version, we have found that the Bogoliubov
transformation eq.(\ref{bogo}) and eq.(\ref{partial}), as well as the transformation of
the vacuum state eq.(\ref{vac}), can all be generated by a operator valued unitary
transformation,
\begin{equation}
U(\sigma,x)=e^{-\frac{1}{4}\sigma\ln(\frac{1+\mathbf{G}}{1-\mathbf{G}})\sigma},
\end{equation}
where $\sigma_i=\theta_i+\theta_i^\dag$, e.g. 
\begin{eqnarray}
\tilde{\theta}&=&U\theta U^{-1}\\
\tilde{\mathcal{H}}&=&U\mathcal{H}U^{-1}\\
|\tilde{0}\rangle&=&U|0\rangle .
\end{eqnarray}
In this language, our model is equivalent to the one analyzed in ref.\cite{mathieu1}
and refs therein by a unitary similarity transformation, hence integrable.


\end{document}